\documentclass{elsart}
\journal{New Astronomy}

\def\astrobj#1{#1}
\def\url#1{{\ttfamily\def\/{/\discretionary{}{}{}}#1}}
\def\bibcode#1{}

\usepackage{hhline,longtable,graphicx,natbib,amsmath}
\bibpunct{(}{)}{;}{a}{}{,}
\begin{document}

\begin{frontmatter}
\title{The orbit of the double-lined Wolf-Rayet Binary HDE~318016 (=~WR~98)}

\author{Roberto C. Gamen\thanksref{fellow}},
\author{Virpi S. Niemela\thanksref{member}}
\address{Facultad de Ciencias Astron\'omicas y Geof\'{\i}sicas, 
       Universidad Nacional de La Plata, Paseo del bosque s/n,
        B1900FWA La Plata, Argentina}
\thanks[fellow]{Fellow of CONICET, Argentina. Visiting Astronomer, CASLEO, San Juan, Argentina. E-mail: rgamen@fcaglp.unlp.edu.ar}
\thanks[member]{Member of Carrera del Investigador, CIC, BA, Argentina. Visiting Astronomer, CTIO, NOAO, operated by AURA, Inc., for NSF. Visiting Astronomer, CASLEO, San Juan, Argentina. E-mail: virpi@fcaglp.unlp.edu.ar}


\begin{abstract}
We present the discovery of OB type absorption lines superimposed to the
emission line spectrum and the first double-lined orbital elements for the 
massive Wolf-Rayet binary HDE~318016 (=WR~98), a spectroscopic binary in
a circular orbit with a period of 47.825 days. The semiamplitudes of
the orbital motion of the emission lines differ from line to line,
indicating mass ratios between 1 and 1.7 for 
$\mathcal{M}_{WR}/\mathcal{M}_{OB}$.
\end{abstract}

\end{frontmatter}

\section{Introduction}

HDE~318016 was discovered to have an emission line spectrum of 
Wolf-Rayet type by \citet{can38}. 
Because both N and C emission lines appear strong in the spectrum it was 
classified as WC7-N6 by \citet{smi68}. This star was included in the
Sixth Catalogue of Galactic Wolf-Rayet Stars \citep{huc81} as WR~98 and
classified as WN7+WC7.
However WR~98 was confirmed to be a single-lined binary with a period of 
47.8 days with N and C emission lines moving in phase \citep{nie91}. 

The optical spectrum of WR~98 has been described by \citet{lun84}.
\citet{con89} proposed the nomeclature WN/WC for WR~98 suggesting that 
this star, along with others with similar emission line spectra where
both N and C lines are observed, has a transition composition between the WN 
and WC subclasses. 

\citet{smi96} classified WR~98 as WN8o/C7, using a new three-dimensio\-nal 
classification scheme, where the ``o'' means that no hydrogen is observed
in the spectrum. 
 Relevant parameters of WR~98 can be found in the recent VIIth Catalogue 
of galactic WR stars \citep{huc01}.

WR~98 is a probable member of the open cluster Trumpler 27 
\citep[e.g.][]{fei00}, and it has also been detected as a non-thermal radio 
source by \citet{abb86}.
Furthermore, WR~98
exhibits random relatively large amplitude optical light variations
($\sim$ 0.1 magnitude in the Johnson V filter), typical of stars with WN8 type
spectrum,
but a periodicity for these variations has not been found \citep{mar98}.

In this work we present a detailed radial velocity analysis of 
optical spectral lines of WR~98 showing it to be a double-lined binary
with high minimum masses.
 
\section{Observations}
\subsection{Photographic spectra}

41 photographic spectrograms were obtained at the Cerro Tololo 
Inter-Ameri\-can Observatory, Chile, between 1980 and 1984.
These spectrograms were secured with the image-tube spectrograph at the 
Cassegrain
focus of the 1-m Yale telescope. All exposures were made on Kodak IIIa-J
emulsion baked in ``forming gas'' ($N_2+H_2$). The spectrograms have a 
reciprocal dispersion of 45 \AA\, mm$^{-1}$. A spectral region from
$\sim$ 3600 to 5000 \AA\, was covered.
Exposure times vary between thirty and ninety minutes, giving Signal to
Noise ($S/N$) ratios $\sim$ 15--30.
A He-Ar lamp spectrum was used as wavelength calibration source.
 
A preliminary report of the binary nature of WR~98 based on these data 
was presented in the IAU Symp. 143 \citep{nie91}.

\subsection{Digital spectra}

28 optical digital spectral images of WR~98 were obtained with the Cassegrain 
Boller \& Chivens (B\&C) and REOSC spectrographs 
attached to the 2.15-m telescope at Complejo Astron\'omico El Leoncito
(CASLEO)\footnote{operated under 
agreement between CONICET, SeCyT, and the Universities of La Plata, 
C\'ordoba and San Juan, Argentina} in San Juan, Argentina, 
between 1997 and 2001.
Nine spectra were secured with the B\&C
spectrograph, in 1997, March, and 1998, February and May. 
A PM~512$\times$512 pixels CCD, with pixel size of 20$\mu$m, was used
as detector. The reciprocal dispersion was $\sim2.3$\,\AA\,pixel$^{-1}$, 
and the wavelength region covered was about $\lambda\lambda$ 3900 -- 4900~\AA.

Nineteen spectra were obtained with the REOSC spectrograph, 
between 1999, March, and 2001, October.
For these spectra a TEK~1024$\times$1024 pixels CCD, 
with pixel size of 24$\mu$m, was used as detector.
The reciprocal dispersion was $\sim1.6$~\AA\,pixel$^{-1}$, and
the observed wavelength region was $\sim \lambda\lambda$ 3850 -- 5450~\AA. 

We used a slit width of 2 arcsecs for all our spectra.
Typical exposure times for the stellar images were between 30 and 40 minutes,
resulting in spectra of signal-to-noise ratio S/N $\sim$~50-100.

He-Ar (or Cu-Ar with REOSC spectrograph) comparison arc images were observed 
at the same telescope position
as the stellar images inmediately after or before the stellar exposures.
Also bias and flat-field frames were obtained each night, as well as
flux and radial velocity standard stars.

All spectra were processed with IRAF 
\footnote{IRAF is distributed by the NOAO, operated by the AURA, Inc,
under cooperative agreement with the NSF, USA.} 
routines at La Plata Observatory. 

\subsection{Determination of radial velocities}

Radial velocities of lines in the photografic spectra of WR~98 were
measured with the Grant oscilloscope
comparator--microphotometer at the Instituto de Astronom\'{\i}a y F\'{\i}sica
del Espacio (IAFE), Buenos Aires, Argentina.

For the present study, selected photografic spectrograms and their 
corresponding calibrations were
digitized with a Grant microphotometer, and calibrated and measured with IRAF
routines.

The emission lines in the spectrum of WR~98 appear approximately
gaussian in shape (cf. Fig~\ref{spect}).
Therefore, for determination of radial velocities in our digital spectra,
we measured central wavelengths of all emission lines 
fitting gaussian functions to the line profiles. 

The journal of the spectroscopic observations with the radial velocity
measurements is presented in Table~\ref{foto},  where
succesive columns quote: the Heliocentric Julian Date of each observation,
the orbital phase,
and the radial velocities for N{\sc iv}, N{\sc v}, C{\sc iii}, He{\sc ii} 
emission lines and H$\delta$ and H$\gamma$ absorption lines, respectively.

{\scriptsize
\begin{longtable}[c]{r r r r r r r r }
\caption[]{Observed heliocentric radial velocities (in km~s$^{-1}$) in the spectra of WR~98}
\label{foto}
\\ \hline
HJD & Phase$^1$ & N\,{\sc iv} em. & N\,{\sc v} em. & C\,{\sc iii} em. & He\,{\sc ii} em. & H$\gamma\,$ abs. & H\,$\beta$ abs.\\
2\,400\,000+
& & 	{\scriptsize $\lambda_0$\,4057.76} &
   	{\scriptsize $\lambda_0$\,4603.73} &
	{\scriptsize $\lambda_0$\,4648.83} &
   	{\scriptsize $\lambda_0$\,4685.68} &
	{\scriptsize $\lambda_0$\,4340.47} &
	{\scriptsize $\lambda_0$\,4861.33} \\
\hline
\endfirsthead
\multicolumn{8}{r}{\em Table~\ref{foto} Continued}\\
\hline
HJD & Phase$^1$ & N\,{\sc iv} em. & N\,{\sc v} em. & C\,{\sc iii} em. & He\,{\sc ii} em. & H$\gamma\,$ abs. & H\,$\beta$ abs.\\
2\,400\,000+
& &     {\scriptsize $\lambda_0$\,4057.76} &
        {\scriptsize $\lambda_0$\,4603.73} &
        {\scriptsize $\lambda_0$\,4648.83} &
        {\scriptsize $\lambda_0$\,4685.68} &
        {\scriptsize $\lambda_0$\,4340.47} &
        {\scriptsize $\lambda_0$\,4861.33} \\
\hline
\endhead
\hline
\endfoot
\endlastfoot
44387.850 & 0.057 &  & 52 & -7 & 27 &  &  \\
44388.780 & 0.076 &  & 10 & -42 & 35 &  &  \\
44390.875 & 0.120 & 60 & 45 & -34 &  &  &  \\
 &  &  &  &  &  &  &  \\
44739.900 & 0.418 & -57 &  & 2 & 51 &  &  \\
44740.830 & 0.438 & -43 & 80 & -38 & 2 &  &  \\
44744.820 & 0.521 &  &  & -79 & -38 & 43 &  \\
 & \multicolumn{1}{l}{ } &  &  &  &  &  &  \\
44895.510 & 0.672 & -118 & -57 & -93 & -74 &  &  \\
44897.520 & 0.714 & -188 & -171 & -102 & -44 & 160 &  \\
44898.520 & 0.735 & -226 & -60 & -113 & -106 &  &  \\
 & \multicolumn{1}{l}{ } &  &  &  &  &  &  \\
45067.880 & 0.276 & -4 &  & 11 & 38 & -47 &  \\
45068.890 & 0.297 & -5 & 87 & 33 & 42 & -128 &  \\
45069.910 & 0.319 &  & 79 & 46 & 88 &  &  \\
45070.910 & 0.339 & -27 & 73 & -9 & 20 & -103 &  \\
 & \multicolumn{1}{l}{ } &  &  &  &  &  &  \\
45123.840 & 0.446 & 88 &  & -10 & 30 &  &  \\
 & \multicolumn{1}{l}{ } &  &  &  &  &  &  \\
45184.640 & 0.718 & -183 &  & -110 & -47 &  &  \\
45186.590 & 0.758 & -190 &  & -142 & -88 &  &  \\
45189.610 & 0.821 &  &  & -96 & -63 &  &  \\
45190.600 & 0.842 &  & -71 & -125 & -67 &  &  \\
45191.620 & 0.863 & -65 & -166 & -65 & -49 &  &  \\
45192.630 & 0.885 & -208 & -109 & -79 & -16 &  &  \\
45196.630 & 0.968 & -51 & -76 & -77 & -24 &  &  \\
45197.670 & 0.990 & -59 & 57 & -78 & -19 &  &  \\
45198.640 & 0.010 & -79 & 80 & -53 & 11 &  &  \\
45199.630 & 0.031 & -29 & 29 & -16 & 9 &  &  \\
 & \multicolumn{1}{l}{ } &  &  &  &  &  &  \\
45248.520 & 0.053 & -25 & 55 & -35 & 25 &  & -221: \\
45249.520 & 0.074 & 36 & 121 & 21 & 43 & -46 &  \\
45250.510 & 0.095 & 83 & 64 & 76 & 86 & -115 &  \\
45251.540 & 0.116 & 93 & 80 & -1 & 68 &  & 12 \\
45252.540 & 0.137 & 9 & 37 & 40 & 66 & -80 & -27 \\
45254.530 & 0.179 & 84 & 131 & -17 & 30 &  &  \\
45255.530 & 0.200 & 58 & -17 & -12 & 23 &  & -72 \\
45256.530 & 0.221 & 139 & 96 & 76 & 113 &  &  \\
45257.540 & 0.242 &  & -26 & 74 & 102 &  &  \\
45258.530 & 0.263 &  & 16 & 21 & 48 &  &  \\
45260.530 & 0.304 & -5 & 99 & 50 & 29 &  & -67 \\
 & \multicolumn{1}{l}{ } &  &  &  &  &  &  \\
45502.750 & 0.369 &  & 113 & -34 & 36 &  &  \\
45506.840 & 0.455 & 68 & -13 & 5 & 62 &  &  \\
45507.830 & 0.475 & 0 & 37 & -19 & 40 &  &  \\
45508.780 & 0.495 &  & -31 & -11 & 24 & 6 & 92 \\
 & \multicolumn{1}{l}{ } &  &  &  &  &  &  \\
45555.670 & 0.476 & 10 & -57 & -39 & 4 &  &  \\
 & \multicolumn{1}{l}{ } &  &  &  &  &  &  \\
45846.750 & 0.562 & 12 & 7 & -61 & -37 &  & 206: \\
 & \multicolumn{1}{l}{ } &  &  &  &  &  &  \\
50536.880 & 0.631 & -120 & -51 & -103 & -41 & 116 & 88 \\
50537.883 & 0.652 & -123 & -133 & -109 & -53 &  &  \\
50538.872 & 0.672 & -124 & -118 & -112 & -34 &  & 118 \\
50539.886 & 0.693 & -125 & -104 & -122 & -78 &  & 39 \\
50540.878 & 0.714 & -176 & -144 & -130 & -55 &  & 86 \\
50541.882 & 0.735 & -130 & -143 & -136 & -59 &  &  \\
50542.877 & 0.756 & -169 & -141 & -100 & -70 &  & 144 \\
 & \multicolumn{1}{l}{ } &  &  &  &  &  &  \\
50858.851 & 0.363 & -38 & 50 & -34 & 26 & -24 & -120 \\
50860.844 & 0.404 & -45 & 5 & -47 & 15 & -47 & -124 \\
 & \multicolumn{1}{l}{ } &  &  &  &  &  &  \\
51261.900 & 0.790 & -93 & -61 & -121 & -41 & \multicolumn{1}{l}{ } &  \\
51261.913 & 0.791 & -125 & \multicolumn{1}{l}{ } & -125 & -38 & \multicolumn{1}{l}{ } &  \\
51264.909 & 0.853 & -106 & -75 & -122 & -49 & \multicolumn{1}{l}{ } &  \\
51266.907 & 0.895 & -96 & -4 & -104 & -31 & \multicolumn{1}{l}{ } &  \\
51267.901 & 0.916 & -37 & -28 & -91 & -34 & \multicolumn{1}{l}{ } & 39 \\
 & \multicolumn{1}{l}{ } &  &  &  &  &  &  \\
51303.768 & 0.666 & -102 & -126 & -105 & -47 & 50 & 135 \\
51304.820 & 0.688 & -192 &  & -112 & -41 &  &  \\
51305.814 & 0.709 & -160 & -136 & -119 & -46 &  & 184 \\
 & \multicolumn{1}{l}{ } &  &  &  &  &  &  \\
51653.819 & 0.985 & -86 & -17 & -68 & 17 &  & -15 \\
51654.833 & 0.006 & -69 & -37 & -56 & 12 &  & 47 \\
 & \multicolumn{1}{l}{ } &  &  &  &  &  &  \\
51806.573 & 0.179 & 134 & 110 & 8 & 110 &  & -191 \\
51807.534 & 0.199 & 87 & 66 & 29 & 81 & -68 & -189 \\
 & \multicolumn{1}{l}{ } &  &  &  &  &  &  \\
52009.863 & 0.430 & -5 & -14 & -27 & 47 & -57 & -74 \\
52010.898 & 0.452 & -21 & -22 & -2 & 42 &  & -7 \\
52011.880 & 0.472 & -15 & -15 & -41 & 37 & 42 &  \\
 & \multicolumn{1}{l}{ } &  &  &  &  &  &  \\
52192.503 & 0.249 & 51 & 36 & 18 & 52 & \multicolumn{1}{l}{ } & -169 \\
52193.496 & 0.270 & 69 & 90 & 29 & 83 & \multicolumn{1}{l}{ } & -68 \\
52194.503 & 0.291 & 66 & 69 & 16 & 64 & \multicolumn{1}{l}{ } & -97 \\
52195.501 & 0.312 & 52 & 73 & 7 & 59 & \multicolumn{1}{l}{ } & -161 \\
\hline
\multicolumn{8}{l}{$^1$ Phases were computed according to 
T$_0$ = 2,445,676.4+47.825E (See below).}\\

\end{longtable}
}

\section{Results and Discussion}
\subsection{The spectrum}

\begin{figure}
\resizebox{\hsize}{!}{\includegraphics{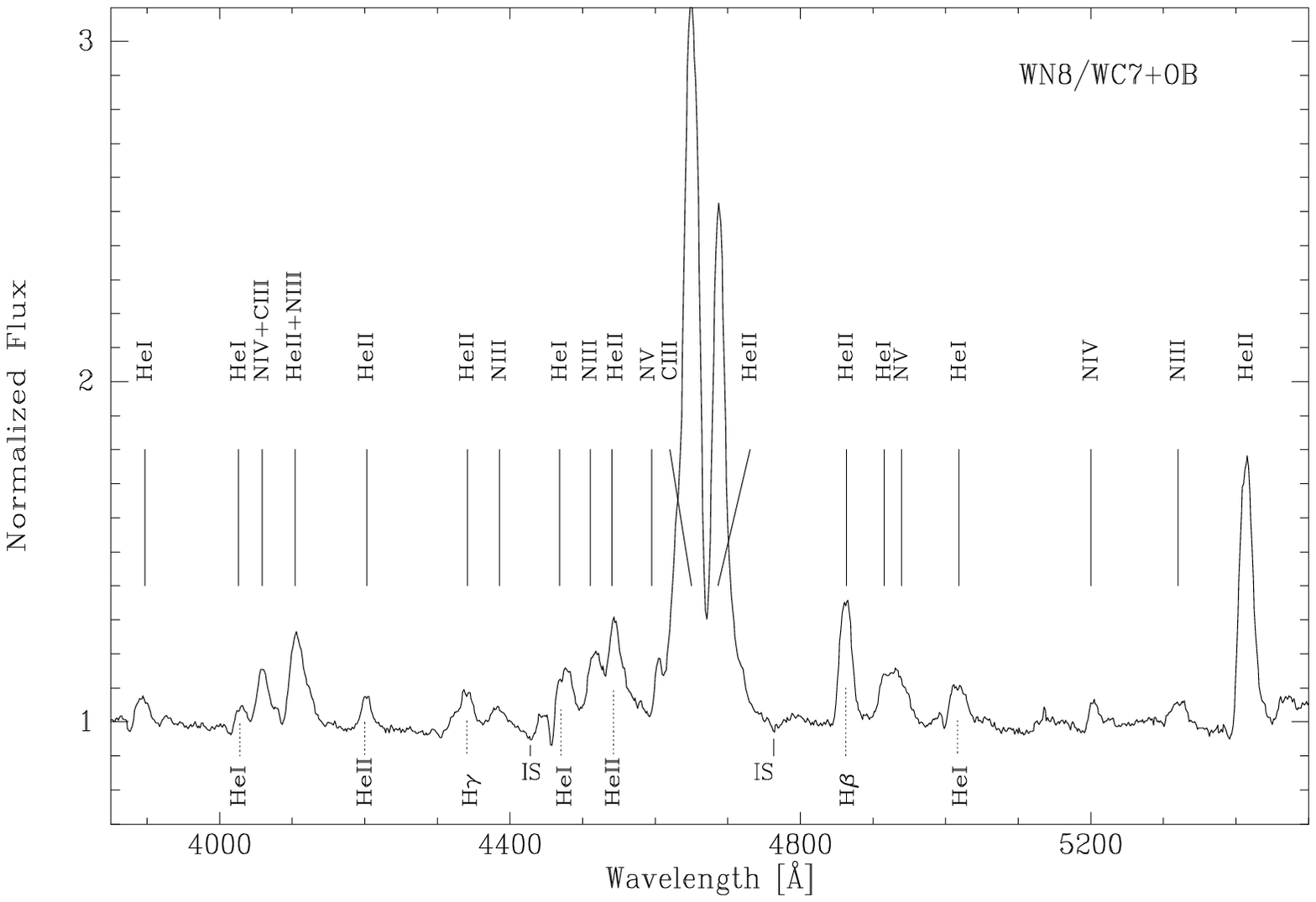}}
\caption {Continuum normalized spectrum of WR\,98 obtained at CASLEO in 2000,
September. Main emission lines are identified above, and companion OB 
absorptions below the spectrum}
\label{spect}
\end{figure}

The blue optical spectrum of WR~98 is illustrated in fig~\ref{spect}, 
with identifications for
main spectral features. As seen in fig~\ref{spect}, the spectrum of 
WR~98 is dominated by emission lines of N{\sc iii}, N{\sc iv}, He{\sc ii},
He{\sc i}, with a strong C{\sc iii} feature at $\lambda$ 4650\AA\,,
consistent with the WN-WC classification originally proposed by \citet{smi68}.
All of He{\sc i} emission lines in our blue spectra 
show P-Cygni profiles, as well as He{\sc ii} 5411\AA\, and 
N{\sc iv} 5203\AA\, (See Fig.~\ref{spect}).
Relative intensities of N{\sc iii}, He{\sc ii}, N{\sc iv}, and C{\sc iii}
lines in our spectra of WR~98 indicate a spectral type 
WN7-8/C.  No hydrogen is detected in the spectrum.
Relative intensities of emission lines in the spectrum of WR~98 show no
appreciable changes in the whole of our dataset spanning 20 years of 
observations.

\begin{figure}
\resizebox{\hsize}{!}{\includegraphics{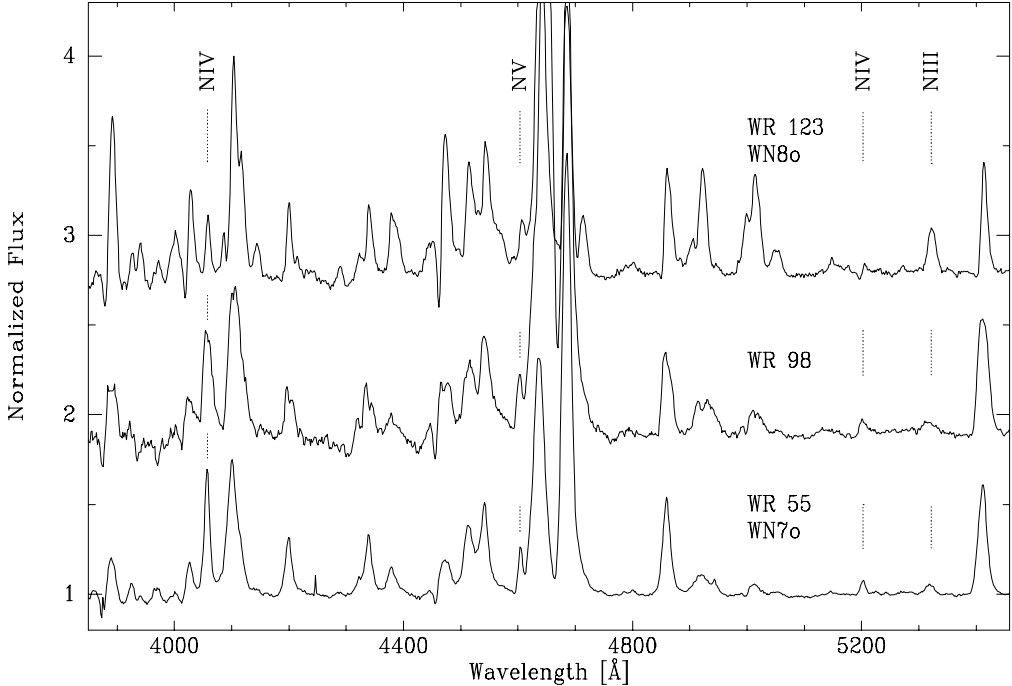}}
\caption {Continuum normalized spectra of WR\,123 (WN8o), WR\,98, and WR\,55
(WN7o). Note the similarity of the WN spectrum of WR~98 with that of WR~55.  
}
\label{spect3}
\end{figure}

We have compared WN the spectrum of WR~98 with that of stars classified as 
WN7o and WN8o \citep{smi96},
namely \astrobj{WR 55} and \astrobj{WR 123}, for which we also have 
digital spectra observed with the same instrumental configuration as WR~98.
This comparison indicates that WR~98 has a higher ionization degree
than the WN8o star, because N{\sc iv} emission lines 
in the spectrum of WR~98 are stronger and N{\sc iii} emission lines are weaker. 
Actually,
the WN spectrum of WR~98 mostly resembles that of WR~55, classified as WN7 by
\citet{smi96}. 
Thus our blue optical spectra of WR~98 would be best described as WN7o/WC.
Fig.~\ref{spect3} shows the spectra of WR\,123, WR\,98, and WR\,55 for
comparison.

Several faint absorption lines were detected upon the WN emission lines
in our spectra of WR~98.
These absorptions are H$\gamma$ and H$\beta$, 
He{\sc i} $\lambda\lambda$ 4026, 4471, 
and 5015 \AA\, and He{\sc ii} $\lambda\lambda$ 4200, and 4540\AA\,.
As will be shown below, these lines belong to an OB companion.
Because He{\sc ii} absorptions appear fainter than He{\sc i}
in the OB spectrum, we presume that the companion 
probably is of spectral type O8-9.
The luminosity class is not possible to determine from our data. 

\subsection{The orbital period}

Previous results \citep{nie91} already have shown the variability of radial
velocities of the emission lines in the spectrum of WR~98.
We searched for periodicities of the radial velocity variations in the 
strongest emission lines in our spectra, namely
He{\sc} $\lambda$ 4686\AA\, and C{\sc iii} $\lambda$ 4648\AA\, 
using algorithms published by \citet{mar80} and \citet{cin95}.
The most probable period obtained by both codes is 47.8 days.

Using this period as initial value, we then calculated orbital solutions
for the radial velocity variations of both emission lines.
We found that the solutions tend to circular orbits,
with the most probable value of the period as $47.825 days \pm 0.005$.
However, other periods close to this value can not be discarded. 

\subsection{The radial velocity orbit}

Orbital elements for each emission line and the H$\gamma$ and H$\beta$ 
absorptions were determined with an improved version of the program 
originally published by \citet{ber68}.
With the present data, the orbits of the 
four emission lines, N{\sc iv}, N{\sc v}, C{\sc iii}, and He{\sc ii} have
negligible eccentricity, thus we have fitted circular orbits for all our 
radial velocities. 
Circular orbital elements for the individual lines are listed 
in Table~\ref{elements}, where $V_{0}$ refers to the center-of-mass 
velocity, $K$, to the semi-amplitude of the radial velocity variations, 
and $T_0$ is the time when the WR star is in the front of the system.

In our orbital fits the three best defined emission lines gave equal $T_0$ 
values within errors, therefore
we adopt as ephemeris for the WR~98 binary system:

\begin{centerline}{T$_0$ = 2,445,676.4+47.825E}\end{centerline}

N{\sc iv}, N{\sc v}, C{\sc iii}, and He{\sc ii} emission lines move in phase,
indicating that they are formed in the same stellar envelope.
This is also the case in the two other known WN/C binaries, 
namely WR~145 (MR~111) and WR~153 (GP Cep) \citep{mas89}.
No detectable phase delays among emission lines are present within the errors 
of the orbital fits.

Semi-amplitudes of the orbital motion of the He{\sc ii} and C{\sc iii} 
emission lines appear lower than those of ionized Nitrogen emission lines. 
This effect is also observed in other WR binaries, e.g. WR~29 \citep{nie00}, 
and may arise if  the He{\sc ii} and C{\sc iii} lines are partly formed 
in the interaction region of the binary components.

\begin{figure}
\resizebox{\hsize}{!}{\includegraphics{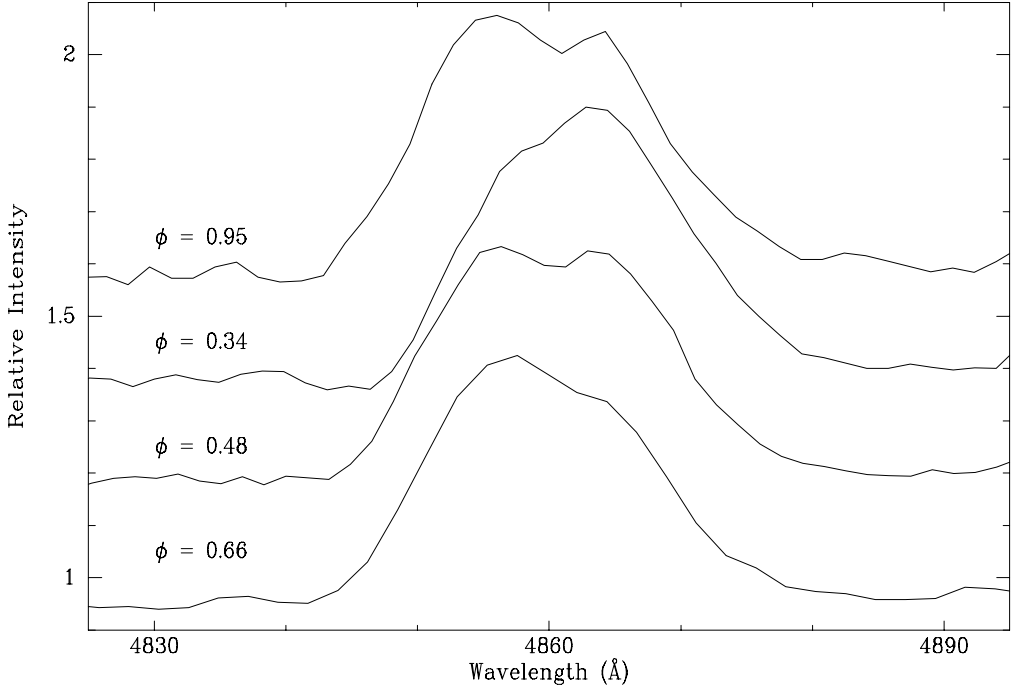}}
\caption {Continuum rectified spectra of the He{\sc ii} 4859\AA\, emission
with the superimposed H$\beta$ absorption
observed during four different phases of the binary system.
Note the antiphased movement of absorption and emission.
The spectra are intensity-shifted for a better presentation.}
\label{comp}
\end{figure}

We measured H$\gamma$ and H$\beta$ absorptions in those spectra of WR~98 were
this was possible. Radial velocities of these hydrogen
lines phased with the binary period move anti-phased with the emission lines,
thus indicating that they belong to an O type companion of the WR 
component in the binary system.
Fig.~\ref{comp} depicts the behaviour of the
H$\beta$ absorption line upon the He{\sc ii} 4859 \AA\, emission in four 
different binary phases, illustrating the antiphased movement of the
absorption and emission lines.

\begin{figure*}
\resizebox{\hsize}{!}{\includegraphics{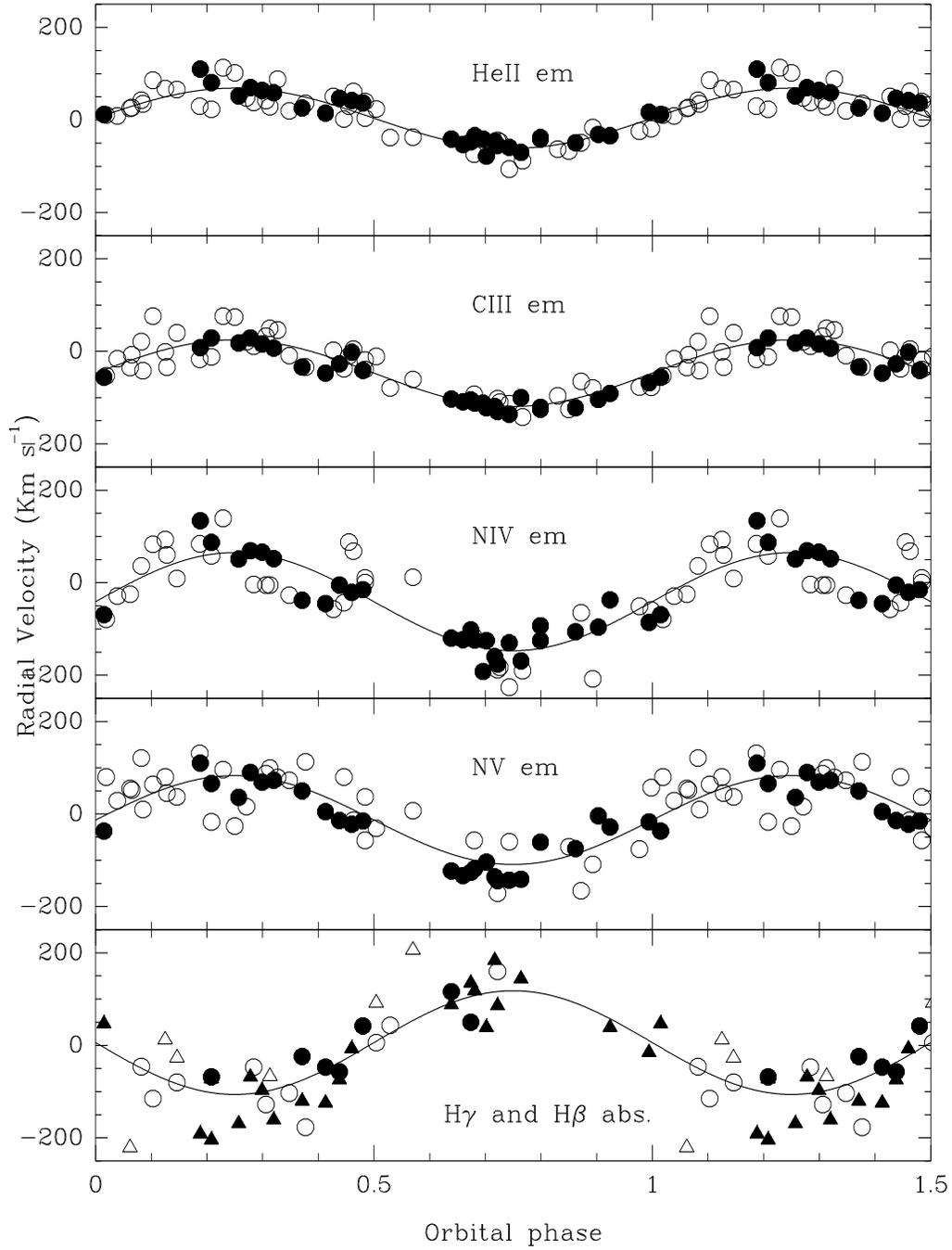}}
\caption {Radial Velocities of He{\sc ii}, C{\sc iii}, N{\sc iv}, and N{\sc v} 
emission lines, and
H$\gamma$ (circles) and H$\beta$ (triangles) absorption lines phased with the
ephemeris 2,445,676.4+47.825E. 
Open symbols represent the photographic data, and solid symbols, the digital
CCD data. Curves are the orbital solutions from Table~\ref{elements}.}
\label{f2}
\end{figure*}

The circular orbital elements for the hydrogen absorption lines are included in
Table~\ref{elements}, and depicted in Fig~\ref{f2}.

The orbital semi-amplitudes of the radial velocity variations of the absorption
lines and of N{\sc iv}, N{\sc v}, He{\sc ii} and C{\sc iii} emission lines 
indicate mass-ratios 
($\mathcal{M}_{WR}/\mathcal{M}_{O}=q$) between 1  and 1.8.
If a monotonic outward decreasing temperature gradient exists in the 
expanding WR envelope, we expect that the highest ionization emission, namely
 N{\sc v}, represents better the orbital motion of the WN/C
component, thus a value of $q \sim 1$ seems more plausible. However, 
taking into account
the rather high uncertainties in the radial velocity values of the absorption
lines, these mass--ratios should not be overinterpreted.
Values of minimum masses and $q$ for
both components are tabulated for each emission line in Table~\ref{elements}.

We also estimated the radii of the critical Roche lobes using the expression
given by \citet{pac71}, which resulted $r_{RL} sin~i = 79.5 R_\odot$ for
both components.  
Each component of the WR~98 binary system seems to be well inside its
critical Roche radius.  

\begin{table*}
\label{elements}
\setlength{\tabcolsep}{0.3mm}
\leavevmode
\caption[]{Circular Orbital Elements of WR~98} 
\begin{tabular}{rl  c rcl c rcl c rcl c rcl c rcl}
\noalign{\smallskip}\hline\noalign{\smallskip}
&&&\multicolumn{15}{c}{WN/C}&&\multicolumn{3}{c}{OB}\\ \hhline{~~~--------------~~---}
\noalign{\smallskip}
\multicolumn{2}{c}{Parameter}     	   & 
& \multicolumn{3}{c}{ ~N\,{\sc  iv} em.~~~}& 
& \multicolumn{3}{c}{ ~N\,{\sc   v} em.~~~}& 
& \multicolumn{3}{c}{ ~C\,{\sc iii} em.~~~}& 
& \multicolumn{3}{c}{~He\,{\sc  ii} em.~~~}& 
& \multicolumn{3}{c}{~absorptions~~~} \\
\noalign{\smallskip} \hline \noalign{\smallskip}
$P$      &  [days] & \multicolumn{19}{c}{47.825 $\pm$ 0.005}\\ 

$V_{0}$ &  [km\,s$^{-1}$] &  & -41 & $\pm$ & 4 &  & -15 & $\pm$ & 3 &  & -47 & $\pm$ & 2 &  & 4 & $\pm$ & 2 &  & 6 & $\pm$ & 15 \\
$K$ &  [km\,s$^{-1}$] &  & 106 & $\pm$ & 6 &  & 109 & $\pm$ & 5 &  & 72 & $\pm$ & 3 &  & 65 & $\pm$ & 3 &  & 112 & $\pm$ & 16 \\
$T_{0}$ &  [HJD]$^{(\ast)}$ &  & 6.4 & $\pm$ & 0.5 &  & 5.5 & $\pm$ & 0.4 &  & 6.5 & $\pm$ & 0.3 &  & 6.3 & $\pm$ & 0.3 &  &   &   &   \\
$a\,\sin i$ &  [R$_\odot$] &  & 100 & $\pm$ & 5 &  & 103 & $\pm$ & 5 &  & 68 & $\pm$ & 3 &  & 61 & $\pm$ & 3 &  & 106 & $\pm$ & 15 \\
$\mathcal{M}_{WN}~sin^{3}i$ & $[\mathcal{M}_\odot]$ &  & 27 & $\pm$ & 10 &  & 28 & $\pm$ & 10 &  & 19 & $\pm$ & 8 &  & 18 & $\pm$ & 7 &  &  &   &  \\
$\mathcal{M}_{OB}~sin^{3}i$ & $[\mathcal{M}_\odot]$ &  & 25 & $\pm$ & 7 &  & 27 & $\pm$ & 7 &  & 12 & $\pm$ & 3 &  & 10 & $\pm$ & 3 &  &  &   &  \\
$q$ &   &  & 1.06 &   &  &  & 1.03 &   &  &  & 1.56 &   &  &  & \multicolumn{1}{c}{1.72} &   &  &  &  &   &  \\
\hline
\multicolumn{22}{l}{$\ast$ HJD 2,445,670+: Time of conjunction, the WN star 
in front of the system.}\\
\end{tabular}
\end{table*}

\subsection{The radial velocity analysis of P-Cygni absorption lines}

\begin{figure}
\resizebox{\hsize}{!}{\includegraphics{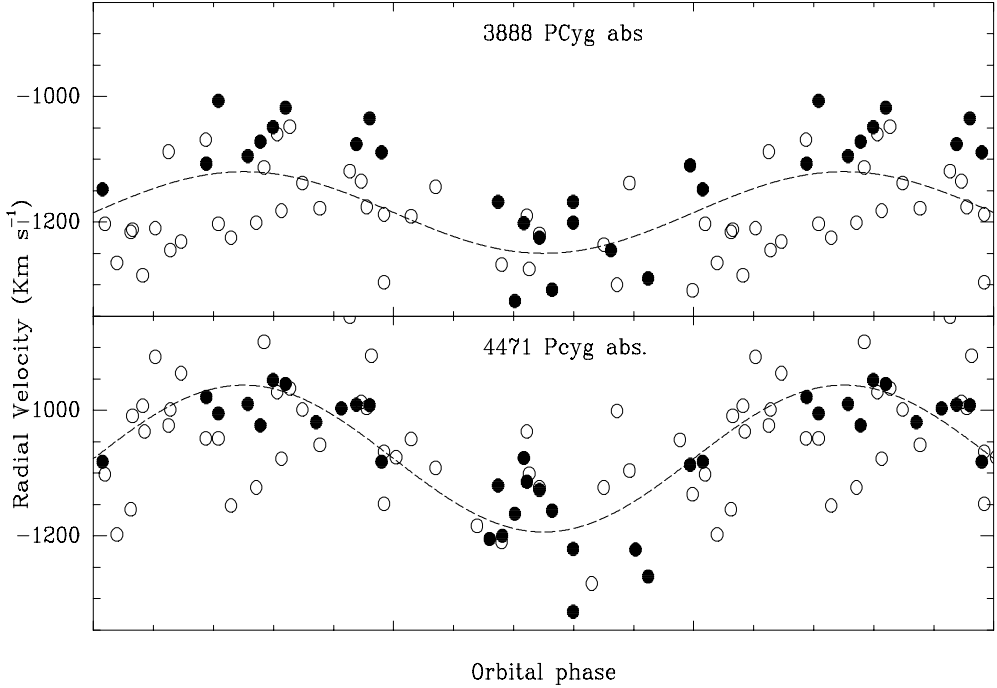}}
\caption {Variations of the radial velocities of two He{\sc i} P-cygni 
absorption lines in the spectrum of WR~98, symbolized and phased with the 
same ephemeris as emission lines in Fig.~\ref{f2}.
Dashed lines represent circular orbital solutions.}
\label{pcyg}
\end{figure}

We analysed the radial velocities of the three strongest P-Cygni absorption 
lines in our spectra of WR~98, namely
He{\sc i} 3888\AA, 4471\AA, and 5015\AA. 
As the He{\sc i} 3888\AA\, absorption presents a rather asymmetric profile, 
we measured the barycenter of this line. Because He{\sc i} 5015\AA\, 
was not observed in the wavelengh range of the photografic
spectra, the radial velocities of this line  were only determined in 
the digital spectra.
The three P-Cygni absorption lines follow the orbital motion of the 
WN/C component of the binary, 
as illustrated in Fig~\ref{pcyg}. 

Circular orbits were fitted to the radial velocity variations of the P--Cyg
absorptions. We obtained similar systemic velocities 
for the three absorption lines:
 $v_0 \sim -1130 \pm 15 km s^{-1}$.
This systemic velocity could be considered as a lower limit of the terminal 
velocity of the WR stellar wind, and in fact the value agrees with the
determination of terminal wind velocity of WR~98 by \citet{een94}.

Semi-amplitudes of the radial velocity variations of the P--Cyg absorptions
gave different values, namely 65 and 104 $km s^{-1}$ for 
He{\sc i} $\lambda\lambda$ 3888, and 4471 and 5015\AA\, respectively. 
The lower semi-amplitude of the orbital motion of He{\sc i} 3888\AA\, 
could be indicating that this line has another component, possibly 
originating in a common expanding envelope surrounding the binary, 
and which is not posible to deblend in our spectra.
Similar behavior of He{\sc i} $\lambda$ 3888\AA\, low-energy metastable 
absorption line is observed in the spectra of other WR binary stars, e.g.
\astrobj{HD 214419} (= WR~155) \citep[See][]{leu83}.

\section {Conclusions}

Our radial velocity analysis of lines in the spectrum of WR~98 based on
a long term spectroscopic database confirms that this star is a binary system 
with a period of 47.825 days. Nitrogen and Carbon emission lines 
in the spectrum move in phase, which indicates that
they are formed in the same stellar envelope. 

We have detected faint absorption lines upon the WR emissions in our spectra. 
These
absorptions belong to an O-type component of the binary, with orbital motion
anti--phased with the WR emissions. 
Thus, WR~98 is one more member of the very limited sample of WR+OB
double-lined binaries, the first one with a WR component which appears to be 
at an intermediate evolutionary phase between the WN and WC stages.

The minimun masses indicated by our radial velocity orbit are quite high. 
Adopting the radial velocity orbit of the N{\sc v} emission as representative
of the motion of the WR component of the binary, we obtain values of minimum
masses of 28 and 27 $M_\odot$ for the WR and OB components, respectively. 
Therefore, the component which at present shows the WR type spectrum, must
 have been very massive during its main sequence stage. This is
in agreement with the estimate of 80 $M_\odot$ for the 
cluster turn-off of Tr 27  \citep{mas01}, of which WR~98 is a probable
member.

A comparison of the WN spectrum of WR~98 with other stars classified as WN8o
(WR~123) and WN7o (WR~55), shows a remarkable resemblance of WR~98 with
the WN7o spectrum, as seen in the Fig.~\ref{comp}.
In our data, the classification of WR~98 as WN7/C appears more likely.

P-Cygni absorption lines of He{\sc i} in our spectra of WR~98 also follow the
orbital motion of the WR component of the binary,
thus these lines are formed in the expanding atmosphere of the WR star.
We determined a systemic velocity of the P--Cyg absorptions 
of $\sim -1130 km~s^{-1}$ for WR~98, which is in good agreement with the value
of the terminal wind velocity published by \citet{een94}.
 
{\bf Acknowledgements}

\medskip
We are indebted to Federico Bareilles for his support in computer facilities.
We thank the Directors and staff of CTIO and CASLEO for the use
of their facilities.
The CCD and data aqcuisition system at CASLEO has been partly financed
by R.M. Rich through U.S. NSF Grant AST-90-15827.
This research was supported in part through IALP-CONICET, Argentina.


\begin{thebibliography}{}

\bibitem[{Abbott} {et~al.}(1986)]{abb86}
{Abbott}, D.~C., {Bieging}, J.~H., {Churchwell}, E., {Torres}, A.~V. 1986, ApJ, 303, 239
\bibcode{1986ApJ...303...239A}

\bibitem[{Bertiau \& Grobben(1968)}]{ber68}
Bertiau, F. \& Grobben, J. 1968, Ric. Astr. Spec. Vat., 8, 1

\bibitem[{{Cannon} \& {Mayall}(1938)}]{can38}
{Cannon}, A.~J. \& {Mayall}, M.~W. 1938, Harvard Observatory Bulletin, 908, 20
\bibcode{1938BHarO.908...20C}

\bibitem[{{Cincotta} {et~al.}(1995){Cincotta}, {Mendez}, \& {Nu\~nez}}]{cin95}
{Cincotta}, P.~M., {Mendez}, M., \& {Nu\~nez}, J.~A. 1995, ApJ, 449, 231
\bibcode{1995ApJ...449..231C}

\bibitem[{{Conti} \& {Massey}(1989)}]{con89}
{Conti}, P.~S., \& {Massey}, P., 1989, ApJ, 337, 251
\bibcode{1989ApJ...337...251C}

\bibitem[{{Eenens} \& {Williams}(1994)}]{een94}
{Eenens}, P.~R.~J. \& {Williams}, P.~M. 1994, MNRAS, 269, 1082
\bibcode{1994MNRAS.269.1082E}

\bibitem[{{Feinstein} {et~al.}(2000){Feinstein}, {Baume}, {Vazquez}, {Niemela},
  \& {Cerruti}}]{fei00}
{Feinstein}, C., {Baume}, G., {Vazquez}, R., {Niemela}, V., \& {Cerruti}, M.~A.
  2000, AJ, 120, 1906
\bibcode{2000AJ....120.1906F}

\bibitem[{{van der Hucht}(2001)}]{huc01}
{van der Hucht}, K.~A. 2001, NewAR, 45, 135
\bibcode{2001NewAR..45..135V}

\bibitem[{{van der Hucht} {et~al.}(1981){van der Hucht}, {Conti}, {Lundstrom},
  \& {Stenholm}}]{huc81}
{van der Hucht}, K., {Conti}, P., {Lundstrom}, I., \& {Stenholm}, B. 1981,
  Space Sci. Rev., 28, 227
\bibcode{1981SSRv...28..227V}

\bibitem[{{Leung} {et~al.}(1983)}]{leu83}
{Leung}, K.-C., {Seggewiss}, W., \& {Moffat}, A.~F.~J. 1983, ApJ, 265, 961
\bibcode{1983ApJ...265..961L}

\bibitem[{{Lundstr\"om} \& {Stenholm}(1984)}]{lun84}
{Lundstr\"om}, I. \& {Stenholm}, B., 1984, A\&ASS, 56, 43
\bibcode{1984A+AS...56...43L}

\bibitem[{{Marchenko} {et~al.}(1998){Marchenko}, {Moffat}, {Eversberg},
  {Morel}, {Hill}, {Tovmassian}, \& {Seggewiss}}]{mar98}
{Marchenko}, S.~V., {Moffat}, A. F.~J., {Eversberg}, T., {Morel}, T., {Hill},
  G.~M., {Tovmassian}, G.~H., \& {Seggewiss}, W. 1998, MNRAS, 294, 642
\bibcode{1998MNRAS.294..642M}

\bibitem[{{Marraco} \& {Muzzio}(1980)}]{mar80}
{Marraco}, H.~G. \& {Muzzio}, J.~C. 1980, PASP, 92, 700
\bibcode{1980PASP...92..700M}

\bibitem[{{Massey \& Grove}(1989)}]{mas89}
Massey, P., and Grove, K. 1989, ApJ, 344, 870

\bibitem[{{Massey et~al.}(2001)}]{mas01}
Massey, P., DeGioia-Eastwood, K., and Waterhouse, E. 2001, AJ, 121, 1050

\bibitem[{{Niemela} \& {Gamen}(2000)}]{nie00}
{Niemela}, V. \& {Gamen}, R. 2000, A\&A, 362, 973
\bibcode{2000A+26A...362..973N}

\bibitem[{{Niemela}(1991)}]{nie91}
{Niemela}, V.~S. 1991, in K.~A.~van der Hucht \& B.~Hidayat (eds.), Wolf-Rayet Stars and Interrelations
  with Other Massive Stars in Galaxies, Proc. IAU Symp. 143, 201
\bibcode{1991IAUS..143..201N}

\bibitem[{{Paczynski}(1971)}]{pac71}
{Paczynski}, B. 1971, Ann. Rev. Astron. Astrophys., 9, 183

\bibitem[{{Smith}(1968)}]{smi68}
{Smith}, L.~F. 1968, MNRAS, 138, 109
\bibcode{1968MNRAS.138..109S}

\bibitem[{{Smith} {et~al.}(1996){Smith}, {Shara}, \& {Moffat}}]{smi96}
{Smith}, L.~F., {Shara}, M.~M., \& {Moffat}, A. F.~J. 1996, MNRAS, 281, 163
\bibcode{1996MNRAS.281..163S}

\end{thebibliography}
\end{document}